\documentclass[10pt,journal]{IEEEtran}

\usepackage{cite}
\usepackage{psfrag}
\usepackage[pdf]{graphicx}
\usepackage[latin1]{inputenc}
\usepackage[T1]{fontenc}
\usepackage{amsmath,amsfonts,amsbsy,amssymb}
\usepackage{mathabx}
\usepackage{mathrsfs}
\usepackage[nolist]{acronym}
\usepackage{tabularx}
\usepackage{amssymb}
\usepackage{amsmath}
\usepackage{graphicx}
\usepackage[latin1]{inputenc}
\usepackage{graphicx}
\usepackage{cite}
\usepackage{multirow}
\usepackage{wasysym}
\usepackage{multirow}
\usepackage{float}
\usepackage{color}

\usepackage[colorlinks=false,urlcolor=black,hidelinks]{hyperref}

\newtheorem{definition}{Definition}

\newtheorem{lemma}{Lemma}
\newtheorem{theorem}{Theorem}
\newtheorem{corollary}{Corollary}
\newtheorem{remark}{Remark}

\newcommand{\mreview}[1]{\textcolor{black}{#1}}
\newcommand{\mreviewB}[1]{\textcolor{black}{#1}}

\begin{document}

\title{Average Error Probability in Wireless Sensor Networks With Imperfect Sensing and Communication for Different Decision Rules}
\author{
	\IEEEauthorblockN{Pedro H. J. Nardelli, Iran Ramezanipour, Hirley Alves, Carlos H.~M. de~Lima and Matti Latva-aho} 
	\thanks{P. H. J. Nardelli, I. Ramezanipour, H. Alves and M. Latva-aho are with the Centre for Wireless Communications (CWC), University of Oulu, Finland. Contact: nardelli@ee.oulu.fi. 
	C. H. M. de Lima is with São Paulo State University (UNESP), São João da Boa Vista, Brazil.
	This work is partly funded by Finnish Academy and CNPq/Brazil (n.490235/2012-3) as part of the joint project SUSTAIN, and by Strategic Research Council/Aka BC-DC project (n.292854).}
}
\maketitle
%
%
\begin{abstract}
This paper presents a framework to evaluate the probability that a decision error event occurs in wireless sensor networks, including sensing and communication errors.
We consider a scenario where sensors need to identify whether a given event has occurred based on its periodic, noisy, observations of a given signal.
Such information about the signal needs to be sent to a fusion center that decides about the actual state at that specific observation time.
The communication links -- single- or multi-hop -- are modeled as binary symmetric channels, which may have different error probabilities.
The decision at the fusion center is based on OR, AND, $K$-OUT-OF-$N$ and MAJORITY Boolean operations on the received signals associated to individual sensor observations.
We derive closed-form equations for the average decision error probability as a function of the system parameters (e.g. number of sensors and hops) and the input signal characterization.
Our analyses show the best decision rule is closely related to the frequency that the observed events occur and the number of sensors.
In our numerical example, we show that the AND rule outperforms MAJORITY if such an event is rare and there is only a handful number of sensors.
Conversely, if there is a large number of sensors or more evenly distributed event occurrences, the MAJORITY is the best choice.
We further show that, while the error probability using the MAJORITY rule asymptotically goes to $0$ with increasing number of sensors, it is also more susceptible to higher  channel error probabilities.
\end{abstract}

\begin{keywords}
Data fusion, distributed detection, wireless sensor networks
\end{keywords}
%

\section{Introduction}
In a recent report, the consulting group McKinsey claims that the Internet of Things (IoT) -- ``sensors and actuators connected by networks to computing systems'' -- have ``a total potential economic impact of \$3.9 trillion to \$11.1 trillion a year by 2025'' \cite{IOT-McKinsey2015}.
Although these numbers and methodology might be questionable, the fact that wireless sensor networks (WSNs), which build the core of IoT, are becoming widespread in almost any possible application area, ranging from energy systems to sleep monitoring \cite{perera2014survey}.

This wide variety of uses implies the nonexistence of a given optimal technology solution that fits all WSNs (e.g. \cite{perera2014survey,Gaj2015}); the requirements of control applications in industries (e.g. \cite{Xu2014}) is far different from in household monitoring (e.g. \cite{kelly2013towards}) and the WSN deployment should reflect these differences.

In this paper, we will focus on general applications with loose reliability requirements so that sensors may have limited computational capabilities.
Our goal here is to understand how to build a more efficient WSN to detect a given event based on an input signal, keeping its design as simple as possible.
We follow here the idea of distributed detection (e.g. \cite{tsitsiklis1993decentralized,Chamberland2007}) so that every sensor estimates the occurrence of such an event and then sends this information to a fusion center, which decides based on the locally processed data; in contrast, by employing a centralized approach, the sensors need to send all their raw observations to the fusion center that then makes a decision.

In signal processing and information theory \mreviewB{\cite{cover}}, the field of distributed detection and estimation, which has a relatively long history (e.g.  \cite{cover,ayanoglu1990optimal,Gubner1993,viswanathan1997distributed,blum1997distributed,sinha2008estimation}), builds an elegant theoretical framework for finding optimal (locally and globally) strategies.
In a series of works dating back to the 80's and 90's, the core theoretical findings of distributed detection were established, as summarized in \cite{viswanathan1997distributed}.
Therein, the authors reviewed the advances on the topic, pointing out three different network topologies: parallel (sensors are not connected to one another), serial (sensors connected in series) and tree (sensors connected following a tree hierarchy).
Different formulations for the detection problem have been then described and their optimal solutions discussed.

For instance, the Neyman-Pearson formulation poses the problem as follows \cite{viswanathan1997distributed}: ``for a prescribed bound on the global probability of false alarm, find (optimum) local and global decision rules that minimize the global probability of miss.''
Other way of formulating the problem is based on Bayesian statistics aiming at minimizing the Bayes risk.
Although they differ in basic aspects, both search ways of optimizing the detection rules based on binary hypothesis tests on the presence of a given signal and they state the likelihood-ratio test as the optimal rule (with different parameters, though).
\mreview{For these cases, the analyses are usually carried out in terms of false positives (also known as type I error, or false-alarm) and false negatives (also known as type II errors, or misdetection).}

Even though these results have been established for decades, there is still a great interest in distributed detection.
Recently, Zhang \textit{et al.} studied in two subsequent papers \cite{zhang2013detection,zhang2012error} the detection error probability in balanced binary relay trees. 
The leaves of the trees (lowest level) are related to the sensors while the root (highest level) is the fusion center that makes the decision.
In between them, relay nodes combine the binary messages sent by their two neighbors at the lower level (either sensors if we consider the second lowest level or other relays for the other levels).
Using the likelihood test at every level, they derive tight bounds of the error probability as a function of the number of sensors considering the binary symmetric channel \cite{zhang2013detection} and erasure channel \cite{zhang2012error}.
\mreview{
	In \cite{Ferrari2006}, the authors assess the probability of decision error of a network with noisy binary symmetric channels between sensors and fusion center and points out that the fusion rule should be optimized with respect to the observed phenomenon.%
}

In \cite{HoTayQue:J15}, the authors combine the idea of robust decentralized detection with social learning.
Among its contributions, \cite{HoTayQue:J15} generalizes to tree and tandem topologies the results of \cite{veeravalli1994minimax}, where a solution to the ``minimax'' robust detection problem (whose objective is to minimize the worst case performance when the probability distribution of the observations is not completely specified) for parallel topology is presented.
In a recent paper \cite{zhang2015event}, the authors combine the advances in spatial models for wireless networks using spatial point process theory (e.g. \cite{nardelli2015throughput} and references therein) and distributed detection.
\mreview{
	In \cite{Ferrari2011}, decentralized detection for clustered sensor networks with hierarchical multi-level fusion is investigated. 
	Authors conclude that the probability of decision error is dominated by the number of decision levels rather than the clustering formation, which renders minimum performance degradation with uniform clustering. 
}
In addition to wireless sensor networks, distributed detection has been applied in the analysis of spectrum sensing strategies for cognitive radio networks as in \cite{umebayashi2012efficient}.

The field of distributed estimation, although similar, has received relatively less attention.
We can cite the following papers as initial works on that problem \cite{ayanoglu1990optimal,Gubner1993,castanon1985distributed}. 
More recently in \cite{Ribeiro2006a,Ribeiro2006b}, the authors studied the case of distributed estimation with  constrained bandwidth of $1$ bit, proposing a class of maximum likelihood estimators.
Other relatively new results can be found in \cite{dogandvzic2006distributed,Leung2015}, while \cite{sinha2008estimation} provides an interesting survey on the topic. 

In this paper, we choose a slightly different way by analyzing a scenario where the quantization and decision rules are given, instead of seeking for optimal detection/estimation schemes \cite{Chamberland2007}.
As previously mentioned, our aim is to assess the average error probability of a WSN with little computational capabilities.
Specifically, our scenario is a set of sensors that periodically measure a given signal to detect whether a given event happens (e.g. if the signal has a value above a certain threshold).
Based on their noisy measurements, the occurrence of such events is mapped into a binary number (e.g. occurrence implies ``$1$'', not occurrence implies ``$0$''), which defines the sensor state.
The sensors need then to send their states to the fusion center via wireless channel, through one or multiple hops.
The relay nodes only forward the information they received.
We assume binary symmetric channels whose associated error probabilities might be different at each level of the multi-hop transmission (but, within the same level, the probabilities are the same).
A decision about the state of the signal is done at the fusion center based on the binary signals related to each sensor.
The decision rules employed by the fusion center are the memoryless Boolean functions OR, AND, $K$-OUT-OF-$N$ and MAJORITY.

Our study targets at answering the following: \textit{Under which conditions a low average error probability can be achieved for the scenario described above?}
For example, the answer for this question for the case of rare events is that a combination of AND decision rule by the fusion center and quantization of the event occurrence with ``$1$'' leads to an average error probability close to the event frequency itself, even when only a handful of sensors are used.
A deeper discussion about this is found later in this paper.
It also important to mention that our work differs from others in the literature as \cite{Ferrari2006, Ferrari2011} by focusing on a single error probability metric rather than cross-over, false-positive or false-negative probabilities.
\mreviewB{More specificaly, the present work generalizes \cite{Ferrari2006} to multi-hop binary symmetric channels and general fusion rules.}
%

In summary, we identify the following main contributions: ($i$) propose an analytic framework  that first breaks the WSN into three phases -- sensing, communication and decision -- to analyze the error events related to each one of them (Section \ref{sec:sys}) to then rebuild the system as a whole to understand how the error propagates through those phases (Section \ref{sec:MainResults}), regardless of the input function; 
($ii$) find a joint error probability which accounts for sensing and communication, while considering that the sensor observations (which are conditionally independent) are subject to Gaussian noise and independent binary symmetric channels with different error probabilities at each hop (Section \ref{sec:sys}); 
($iii$) derive, in closed-form, the average error probability for the OR, AND, $K$-OUT-OF-$N$ and MAJORITY (Section \ref{sec:MainResults}), showing that MAJORITY rule can asymptotically  reach a $0$-error probability with the number of sensors; ($iv$) show that the performance of OR and AND rules depends on the frequency of the event under analysis, while the other two do not -- this fact implies that, when a limited number of sensors is considered, OR or AND rule (depending on the quantization mapping) can outperform the MAJORITY;
($v$) exemplifying the analysis with numerical results (Section \ref{sec:NumericalAnalysis}) to illustrate our findings, providing the basis for our final discussions and possible extensions (Section \ref{sec:Discussions-final-remarks}).

\section{System description}
\label{sec:sys}
\begin{table}[!b]
	\centering
	\caption{\mreview{Notation summary}}
	\label{tab:notation}	
	\centering
	\mreview{
	\begin{tabular}{|c|p{.7\columnwidth}|}
		\hline
		\textbf{Notation}	&	\textbf{Description}		\\ \hline \hline
		$\mathcal{N}$ & set of sensors \\ \hline
		$N$ & number of sensors in set $\mathcal{N}$ \\ \hline
		$M$ & number of communication hops \\ \hline
		$\eta$ & total number of samples \\ \hline
		$x(t)$ & continuous signal as a function of time $t$		\\ \hline
		$x_\mathrm{th}$ & signal threshold \\ \hline	
		$\theta[n]$ & sensor state at time $t_n$ related to the $n$th measurement\\ \hline
		$y_i[n]$ & sensor $i$ estimation about the $n$th measurement\\ \hline
		$\mathcal{E}$ & event related to $x(t)$ \\ \hline
		$s_i[n]$ & received state at the fusion center from sensor $i$
		\\ \hline
		$g(\cdot)$ & decision function \\ \hline
		$\hat{\theta}[n]$ & estimated stated at the fusion center\\ \hline
		$n_i(t_n)$ & additive noise associated with sensor $i$ at $t_n$ \\ \hline
		$\mu$ & noise mean \\ \hline
		$\sigma^2$ & noise variance \\ \hline
		$S$ & Boolean variable associated with event $\mathcal{E}$  \\ \hline
		$f_S$ & frequency of a given state $Ss$ \\ \hline
		$S_{i,j}$ & state of sensor $i$ in hop $j$ \\ \hline
		$p_j$ & error probability in the communication hop $j$ \\ \hline
		$\mathcal{C}_M$ & set of all possible error events for $M$-hop link \\ \hline
	\end{tabular}
	}
\end{table}

Let us assume a network composed by a set $\mathcal{N} = \{1,...,N\}$ of sensors that monitor a continuous signal $x(t)$, where $t \in \mathbb{R}^+$ and $x:\mathbb{R}^+\rightarrow\mathbb{R}$, to estimate whether a given event $\mathcal{E}$ related to $x(t)$ happens and send this information to a fusion center. To avoid confusion, we summarize the key notations in Table~\ref{tab:notation}.
Assuming that the sensors make synchronous and periodic measurements in predetermined instants $t_n = n\tau$ with $n$ being a natural number and $\tau \in \mathbb{R}^+$, we can define a function $\theta[n]$ with $\theta: \mathbb{N}\rightarrow\{0,1\}$ that indicates if $\mathcal{E}$ occurs at time $t_n$.
Hereafter we refer to $\theta[n]$ as the system state at time $t_n$. 

The sensors' estimation of $\theta[n]$ is, however, imperfect.
For each sensor $i \in \mathcal{N}$, we define a function $y_i[n]$ with $y_i: \mathbb{N}\rightarrow\{0,1\}$ that represents the estimation about  $\mathcal{E}$ from its individual noisy version of $x(t)$.
If a sensing error at sensor $i$ happens at $t_n$, then $y_i[n] \neq \theta[n]$; otherwise $y_i[n] = \theta[n]$.

After the sensing phase, the sensors need then to forward $y_i[n]$ to a remotely located fusion center, which will process the received information to determine whether $\mathcal{E}$ has indeed occurred.
Each sensor $i \in \mathcal{N}$ sends its state $y_i[n]$ through independent communication channels that are also subject to errors.
Let $s_i[n]$ with $s_i: \mathbb{N}\rightarrow\{0,1\}$ be the state related to sensor $i$ that is received by the fusion center after passing through the radio links, which can be composed by only one hop or multiple hops where relay nodes forward the received information.
If an error occurs in the link related to sensor $i$'s $n$th measurement, then $s_i[n] \neq y_i[n]$; if not, $s_i[n] = y_i[n]$.

From the signals $s_i[n]$, the fusion center needs to decide whether event $\mathcal{E}$ happened at $t_n$.
Let $g(s_1[n],...,s_N[n])$, with $g: \{0,1\}^N \rightarrow\{0,1\}$, denote the Boolean function that estimates the state $\theta[n]$ by the fusion center so that the estimated state $\hat{\theta}[n]$ related to $t_n$ is given by $\hat{\theta}[n] = g(s_1[n],...,s_N[n])$.
A decision error occurs whenever $\hat{\theta}[n] \neq \theta[n]$.
The average error probability $P_\textup{e}$ of the whole process is given by:
%
\begin{equation}
P_\textup{e}=\dfrac{1}{\eta}\;\sum\limits_{n=0}^{\eta-1} \mathrm{Pr}\left[\hat{\theta}[n] \neq \theta[n]\right],
\label{eq:AveErrorProb}
\end{equation}
where the average is taken over the samples $n \in \{0,1,...,\eta-1 \}$ related to a time window from $t_0=0$ and $t_{\eta-1} = T$.

\begin{figure}[!t]%
	\centering
	\includegraphics[width=0.7\columnwidth]{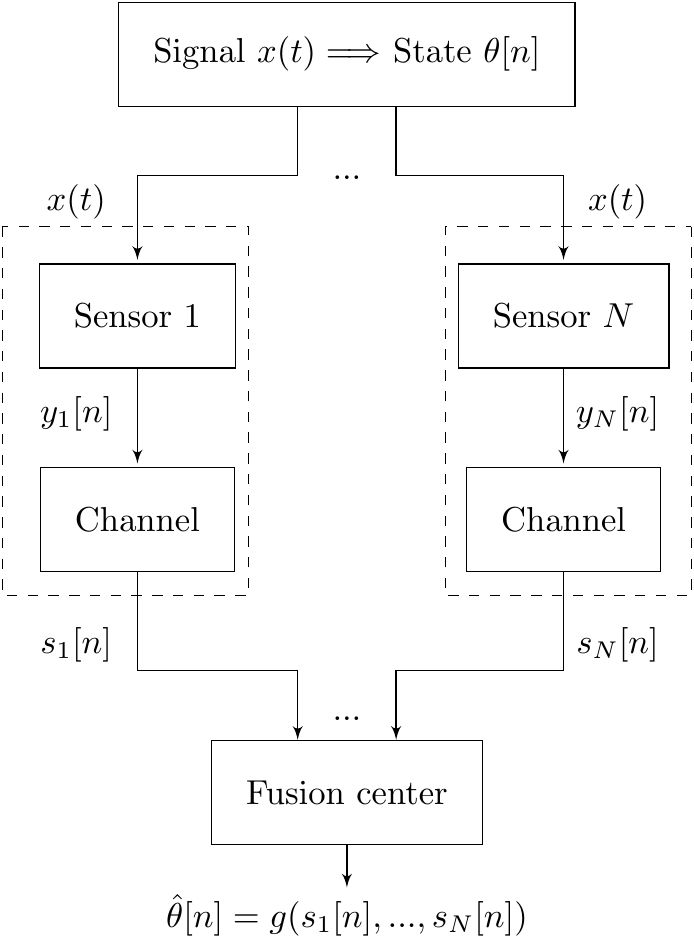}
	%
	\caption{Illustrative figure of the scenario under analysis. Sensors monitor a given signal $x(t)$ in order to determine the binary state $\theta[n]$ at time $t_n$. Each one of the $N$ sensors in the network needs to send its state to a fusion center (control unit) that remotely decides the state $\hat{\theta}[n]$. In its way to the fusion center, errors may happen either in sensing ($y_i[n] \neq \theta[n]$ with $i=1,.., N$) or in communicating ($s_i[n] \neq y_i[n]$). The dashed rectangle identifies where  the error events may happen. Our goal is to find the expected error probability $P_\textup{e}$ given by \eqref{eq:AveErrorProb} and then compare different design options.}
	\label{fig:scheme}
	\vspace{-4mm}
\end{figure}

Our goal here is to analyze different design options for the sensor network and the fusion center's decision function to improve the estimation reliability, evaluated by the probability that  $\hat{\theta}[n] \neq \theta[n]$.
Fig. \ref{fig:scheme} illustrates the scenario under analysis.
For instance, $x(t)$ may represent the temperature of an industrial plant that requires temperatures below a given threshold $x_\textup{th}$ to guarantee its safe operation.
The event $\mathcal{E}$ can be then associated with an emergency where $x(t_n) > x_\textup{th}$.
Given the signal $x(t)$, the threshold $x_\textup{th}$ and the number of sensors $N$, we need to find the most suitable design for the quantization function $\theta$ (i.e. define  if the event $x(t) > x_\textup{th}$ is associated with $\theta = 0$ or $\theta = 1$) and the decision rule $g$ (OR, AND, $K$-OUT-OF-$N$, MAJORITY) at the fusion center.

Next we will focus our attention on the errors in the sensing procedure and in the communication links, which are identified by the dashed rectangle in Fig. \ref{fig:scheme}, and how they affect the decisions done by the fusion center.

\vspace{-3mm}

\subsection{Sensing error}
\label{subsec:SensErrors}
\mreview{
Let us denote $x_i(t_n)$ the version of $x(t)$ observed by sensor $i \in \mathcal{N}$.
The value of $x_i(t_n)$ will be then used to define $y_i[n]$.
Then, we can define the probability $P(x_i(t_n) \leq x_\textup{th})$ that the event $x_i(t_n) \leq x_\textup{th}$, and its complement $P(x_i(t_n) > x_\textup{th})$.
}

\begin{remark}
The sensing error probability is dependent on the input signal $x(t)$ such that an error in the sensing procedure $y_i[n] \neq \theta[n]$ occurs in two situations: (a) $x(t_n) > x_\textup{th}$ and  $x_i(t_n) \leq x_\textup{th}$, or  (b) $x(t_n) \leq x_\textup{th}$ and  $x_i(t_n) > x_\textup{th}$.
The error probability is then related to the frequency that $x(t_n)$ is above or below the threshold $x_\textup{th}$, which is captured by how many times $\theta[n]=0$ or $\theta[n]=1$ for $n=0,...,\eta-1$.
\end{remark}

Let us consider  $x(t_n) \leq x_\textup{th}$ is associated to the state $\theta[n]=S$ where $S \in \{0,1\}$ and $x(t_n) > x_\textup{th}$ is associated to the state $\theta[n]=\bar{S}$, where $\bar{S}$ denotes the complement of $S$.
Then, we have the following definition.

\mreview{
\begin{definition}
\label{def:defining-theta}
Recalling that $\eta$ denotes the total number of samples within an arbitrary interval defined by $t_0=0$ and $t_{\eta-1}=T$.
We define $\eta_{S} \triangleq \eta_{x(t_n) \leq x_\textup{th}}$ and $\eta_{\bar{S}} \triangleq \eta_{x(t_n) > x_\textup{th}}$ as the number of samples related to states $S$ and $\bar{S}$ in such interval, respectively.
In this case, $\eta_{S} + \eta_{\bar{S}} = \eta$.
%
%
The frequency $f_{S} = f_{x(t_n) \leq x_\textup{th}}$ that the state $S$ appears between $t_0$ and $t_{\eta-1}$ is
%
%
$f_{S} = \eta_{S} /\left({\eta_{S} + \eta_{\bar{S}}}\right)$
\label{eq:freq-S}
%
Similarly, the frequency $f_{\bar{S}} = f_{x(t_n) > x_\textup{th}}$ is
%
%
$f_{\bar{S}} = {\eta_{\bar{S}}} /\left({\eta_{S} + \eta_{\bar{S}}} \right) $. 
%
%
\end{definition}
}

\subsection{Communication errors}
\label{subsec:CommErrors}
\begin{figure}[!t]%
	\centering
	\includegraphics[width=.85\columnwidth]{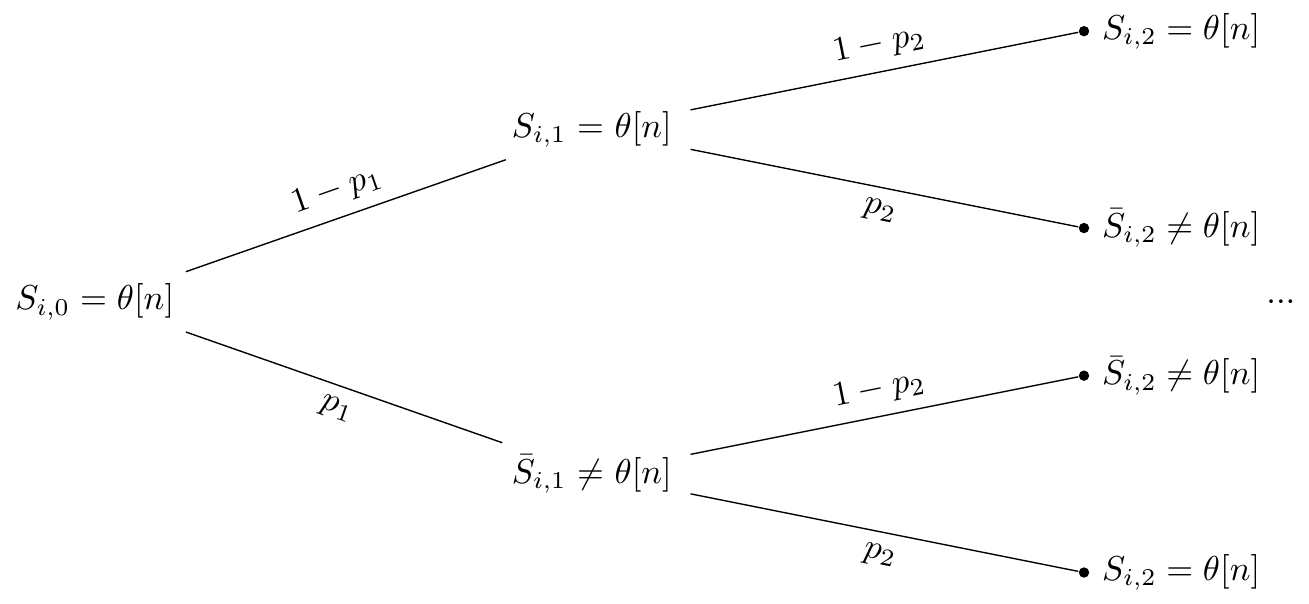}
	%
	\caption{State probability tree for each sensor $i=1,...,N$ considering communication error probabilities $p_j$ with $j=1,...,M$ where $M$ is the number of hops between the sensors and the fusion center.  The initial state is $S_{i,0} = y_i[n]$ and the final state is $S_{i,M}$.  The state $S_{i,j} \in \{0,1\}$ and  $\bar{S}_{i,j}$ denotes its complement.}
	\label{fig:error-tree}
	\vspace{-4mm}
\end{figure}

Let $S_{i,0}[n] = y_i[n]$ be the signal sent by sensor $i$ and $S_{i,j}[n]$ be the state at the $j$th level with $j=1,...,M$  communication hop.
At every hop, relay nodes forward their state $S_{i,j}[n]$ to the next one.
The probability tree of the state of sensor $i$ is presented in Fig. \ref{fig:error-tree}, where the initial state $S_{i,0}[n]$ is equal to the state $y_i[n]$ after the sensing procedure and $S_{i,j}[n]$ is the state at the $j$th level.

\begin{figure}[!t]%
	\centering
	\includegraphics[width=0.7\columnwidth]{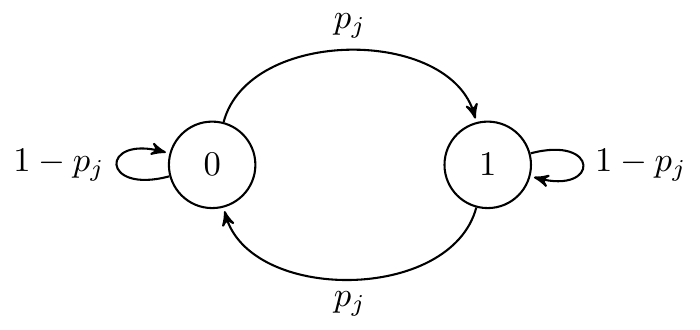}
	%
	\caption{Transition diagram between the two possible states ``0'' and ``1'' when the Binary Symmetric Channel is assumed. A change only occurs with probability $p_j$ with $j=1,...,M$, which is associated with the $j$th level error probability such that  $S_{i,j} = S_{i,j-1}$ with probability $1-p_j$ and $S_{i,j} \neq S_{i,j-1}$ with probability $p_j$.}
	\label{fig:states-prob}
	\vspace{-4mm}
\end{figure}

We assume here a binary symmetric channel where the output is different from the input with probability $p_j$ for the $j$th  level of the tree.
We assume that the error events are independent at each level and time-steps $t_n$, which allows for dropping the index $n$.
Then, the state $S_{i,j} = S_{i,j-1}$ with probability $1-p_j$ and $S_{i,j} \neq S_{i,j-1}$ with probability $p_j$.
Fig. \ref{fig:states-prob} represents the state transition diagram of this channel.
\mreview{
\begin{definition}
\label{lemma:OddErrors}
A communication error event $s_i[n] \neq y_i[n]$ related to sensor $i$ and $t_n$ occurs after $M$ hops if an odd number of errors  $S_{i,j} \neq S_{i,j-1}$ had happened for $j = 1,...,M$.
\end{definition}
}

\begin{definition}
\label{def:ErrorEventsSet}
Let us define the set of all possible error events for an $M$-hop link as $\mathcal{C}_{M} = \left\{\{ \; \}, \{ 1 \},\{ 2 \},...,\{1,2\},\{1,3\},..., \{1,2,...,M\} \right\}$, which  contains $2^M$ elements that refer to the index $j=1,...,M$.
Let $\mathcal{C}_{M,\textup{odd}} \subset \mathcal{C}_{M}$ denote the subset of index with odd cardinality, which is composed by $2^{M-1}$ elements.
The subset $\mathcal{C}_{M,\textup{odd}}^k \subset \mathcal{C}_{M,\textup{odd}}$ with $k = 1,...,2^{M-1}$ denotes each one of the $k$ subsets of $\mathcal{C}_{M,\textup{odd}}$ such that $\bigcup \mathcal{C}_{M,\textup{odd}}^k = \mathcal{C}_{M,\textup{odd}}$.
The complement $\bar{\mathcal{C}}_{M,\textup{odd}}^k$ is defined in relation to $\{1,2,...,M\}$ such that $ \mathcal{C}_{M,\textup{odd}}^k + \bar{\mathcal{C}}_{M,\textup{odd}}^k= \{1,2,...,M\}$.
\end{definition}

Let us now exemplify the construction of those sets when $M=3$.
Following the procedure presented in Definition \ref{def:ErrorEventsSet}, $\mathcal{C}_{3} = \left\{\{ \; \}, \{ 1 \},\{ 2 \},\{3\},\{1,2\},\{1,3\},\{2,3\}, \{1,2,3\} \right\}$, which has eight $(2^3)$ elements.
The subset is then $\mathcal{C}_{3,\textup{odd}} = \left\{ \{ 1 \},\{ 2 \},\{ 3 \},\{1,2,3\}\right\}$, having then four $(2^2)$ elements.
Using these sets, we have $\mathcal{C}_{3,\textup{odd}}^1 = \{ 1 \}$ and $\bar{\mathcal{C}}_{3,\textup{odd}}^1 = \{ 2,3 \}$, $\mathcal{C}_{3,\textup{odd}}^2 = \{ 2 \}$ and $\bar{\mathcal{C}}_{3,\textup{odd}}^2 = \{ 1,3 \}$, $\mathcal{C}_{3,\textup{odd}}^3 = \{ 3 \}$ and $\bar{\mathcal{C}}_{3,\textup{odd}}^3 = \{ 1,2 \}$, and $\mathcal{C}_{3,\textup{odd}}^4 = \{ 1,2,3 \}$ and $\bar{\mathcal{C}}_{3,\textup{odd}}^4 = \{ \; \}$.

\begin{theorem}
\label{theorem:ErrorProbSensor-i}
The communication error probability $P(s_i[n] \neq y_i[n])$ related to sensor $i$ and $t_n$  over $M$ hops is given by:
\begin{equation}
P(s_i[n] \neq y_i[n]) =  \sum\limits_{k=1}^{2^{M-1}} \left( \prod\limits_{i\in\mathcal{C}_{M,\textup{odd}}^k} p_i \right) \left(  \prod\limits_{j\in\bar{\mathcal{C}}_{M,\textup{odd}}^k}(1- p_j)\right).
\label{eq:ErrorProbSensor-i}
\end{equation}
\end{theorem}

\begin{IEEEproof}
From Lemma \ref{lemma:OddErrors} we know  an error event  $s_i[n] \neq y_i[n]$  occurs whenever an odd number of error events $S_{i,j} \neq S_{i,j-1}$ happens for $j = 1,...,M$.
To compute the probability of such events, we need to use the law of total probability knowing that the error events at each level and in different measurement instants $t_n$ are independent.
Using Definition \ref{def:ErrorEventsSet} to characterize the subsets containing the odd number of error events, we obtain \eqref{eq:ErrorProbSensor-i}.
\end{IEEEproof}
\mreview{
\begin{remark}
\label{cor:symmetric-BSC-cascade}
If $p_j = p$ for all $j=1,...,M$, then 
\begin{equation}
P(s_i[n] \neq y_i[n]) = \dfrac{1-(1-2p)^M}{2},
\label{eq:ErrorProbSensor-BSCwith-same-prob}
\end{equation}
which is known result for cascade of binary symmetric channels \mreviewB{\cite{cover}}. 
\end{remark}
}

\subsection{Decision function}

\begin{lemma}
\label{lemma:DecisionErrors}
A decision error event $\hat{\theta}[n] \neq \theta[n]$ at fusion center depends on the actual state $\theta[n]$ and the decision function  $\hat{\theta}[n] = g(s_1[n],...,s_N[n])$ as follows.
\begin{itemize}
\item OR: If $\theta[n] = 0$, then at least one signal $s_i[n] = 1$ with $i \in \mathcal{N}$ leads to $\hat{\theta}[n]=1 \neq \theta[n]$. If $\theta[n] = 1$, then an error event $\hat{\theta}[n]=0 \neq \theta[n]$ only occurs when all signals $s_i[n] = 0$.
\item AND: If $\theta[n] = 0$, then an error event $\hat{\theta}[n]=1 \neq \theta[n]$ only occurs when all signals $s_i[n] = 1$. If $\theta[n] = 1$, then at least one signal $s_i[n] = 0$ with $i \in \mathcal{N}$ leads to $\hat{\theta}[n]=0 \neq \theta[n]$.
\item  $K$-OUT-OF-$N$: If $\theta[n] = 0$, an error event $\hat{\theta}[n] = 1\neq \theta[n]$ occurs when at least $K$ out of $N$ signals $s_i[n] = 1$. If $\theta[n] = 1$, an error event $\hat{\theta}[n] = 0 \neq \theta[n]$ occurs when at least $K$ out of $N$ signals $s_i[n] = 0$. If $N$ is an even number and $K = N/2$, the event where $N/2$ signals are $s_i[n] = 0$ and the other $N/2$ are $s_i[n] = 1$ may occur and the decision will be randomized such that $\hat{\theta}[n]=0$ or $\hat{\theta}[n]=1$ with $50$\% of chance.
\item MAJORITY: This rule is a specific case of $K$-OUT-OF-$N$ when $K=\lceil N/2 \rceil$, where $\lceil{a}\rceil: \; \mathbb{R} \rightarrow \mathbb{Z}$ represents the ceiling function such that it maps the real number $a$ to its smallest following integer. 
\end{itemize}
\end{lemma}

\begin{corollary}
\label{cor:AND-OR-as-K-out-of-N}
The AND and OR decision functions are conditional versions of $K$-OUT-OF-$N$ rule as follows.
\begin{itemize}
\item OR:  If $\theta[n] = 0$, the error event is equivalent to $1$-OUT-OF-$N$ rule. If $\theta[n] = 1$, it is equivalent to $N$-OUT-OF-$N$.
\item AND:  If $\theta[n] = 0$, the error event is equivalent to $N$-OUT-OF-$N$ rule. If $\theta[n] = 1$, it is equivalent to $1$-OUT-OF-$N$.
\end{itemize}
\end{corollary}

\section{Main results}
\label{sec:MainResults}
In this section we combine the results previously presented to derive the main results of this paper, namely the average error probability given in \eqref{eq:AveErrorProb} for the OR, AND, $K$-OUT-OF-$N$ and MAJORITY decision rules as a function of the number of sensors $N$, number of hops $M$ and other system variables.

\begin{lemma}
\label{lem:prob-K-out-of-N}
The probability $P_{\textup{e},K,S}[n]$ that $K$ out of $N$ sensors experience error events $s_i[n] \neq \theta[n]$ at $x(t_n)$ for $x(t_n) \leq x_\textup{th}$ (i.e. $\theta[n]=S$) is:
\begin{equation}
P_{\textup{e},K,S}[n] = \binom{N}{K} (P_{S,\bar{S}}[n])^K   (P_{S,S}[n])^{N-K}, 
\label{eq:error-prob-K-out-of-N-S=0}
\end{equation}
where the probabilities $P_{S,\bar{S}}[n] = P(x(t_n)\leq x_\textup{th} ) P(s_i[n] \neq y_i[n]) + P(x(t_n) > x_\textup{th})  P(s_i[n] = y_i[n])$ and $P_{S,S}[n]= 1 - P_{S,\bar{S}}[n]$.

Similarly, the probability $P_{\textup{e},K,\bar{S}}[n]$ that $K$ out of $N$ sensors experience error events $s_i[n] \neq \theta[n]$ at $x(t_n)$ for $x(t_n) > x_\textup{th}$ (i.e. $\theta[n]=\bar{S}$) is:
\begin{equation}
P_{\textup{e},K,\bar{S}}[n] = \binom{N}{K} (P_{\bar{S}, S}[n])^K   (P_{\bar{S},\bar{S}}[n])^{N-K}
\label{eq:error-prob-K-out-of-N-bar-S=1}
\end{equation}
where the probabilities $P_{\bar{S},S}[n] = P(x(t_n) > x_\textup{th})  P(s_i[n] \neq y_i[n]) + P(x(t_n) \leq x_\textup{th})  P(s_i[n] = y_i[n])$ and  $P_{\bar{S},\bar{S}}[n] = 1 - P_{\bar{S},S}[n]$.
\end{lemma}

In other words, $P_{\textup{e},K,S}[n]$ represents the probability that the  signals related to $K$ sensors are in state $\bar{S}$ and $N-K$ sensors are in state $S$ when arriving at the fusion center, given that $\theta[n] = S$ (i.e. $x(t_n) \leq x_\textup{th}$).

\begin{definition}
\label{def:average}
Let $\mathrm{Av}( \cdot, \cdot )$ be the average operator such that 
\begin{equation}
\mathrm{Av}(v[n],\mathcal{V}) = \dfrac{1}{\#(v[n])}\sum\limits_{n \in \mathcal{V}} v[n],
\label{eq:AveOp}
\end{equation}
where $v[n]$ is a list of numbers, $\#(v[n])$ represents its cardinality and $\mathcal{V}$ is the set containing the indexes related to each one of the $\#(v[n])$ elements of $v[n]$.
\end{definition}

\begin{theorem}
\label{theorem:ErrorProbDecision}
If the state $S = 0$ (i.e. $\theta[n] = 0$ if $x(t_n) \leq x_\textup{th}$; refer to Definition \ref{def:defining-theta}) the expected decision error probability $P_{\textup{e, OR},0}$ introduced in \eqref{eq:AveErrorProb} using OR decision rule and $N$ sensors is
\begin{equation}
P_{\textup{e, OR}, 0} = f_{0}  \mathrm{Av}\left(1 - P_{\textup{e},N,0}[n] , \mathcal{S}_{0}\right)   + f_1 \mathrm{Av}\left(P_{\textup{e},N,1}[n] , \mathcal{S}_{1}\right),
\label{eq:ErrorProbOR}
\end{equation}
where $f_0$ and $f_1$ are given in Definition \ref{def:defining-theta}, and $\mathcal{S}_{0}$ and $\mathcal{S}_{1}$ denote the set containing the indexes related to $S=0$ and $S=1$.

Similarly, the probability $P_{\textup{e, AND},0}$ using AND decision is:
\begin{equation}
P_{\textup{e, AND}, 0} =  f_{0}  \mathrm{Av}\left(P_{\textup{e},N,0}[n], \mathcal{S}_{0}\right)   + f_1 \mathrm{Av}\left(1-P_{\textup{e},N,1}[n] , \mathcal{S}_{1}\right).
\label{eq:ErrorProbAND}
\end{equation}

For the $K$-OUT-OF-$N$ rule except when both $N$ is even and $K = N/2$, the probability $P_{\textup{e}, K, 0} $ is:
\begin{equation}
P_{\textup{e}, K, 0} =  \sum\limits_{k=K}^{N}  f_0 \mathrm{Av}\left(P_{\textup{e},k,0}[n], \mathcal{S}_{0}\right)   + f_1 \mathrm{Av}\left(P_{\textup{e},k,1}[n] , \mathcal{S}_{1}\right).
\label{eq:ErrorProbK-out-of-N}
\end{equation}

For MAJORITY and $N$ odd, the probability $P_{\textup{e, MAJ},0}^\textup{ odd} $ is:
\begin{equation}
P_{\textup{e, MAJ}, 0}^\textup{ odd} =  \sum\limits_{k=\lceil N/2 \rceil}^{N} \hspace{-1ex} f_0 \mathrm{Av}\left(P_{\textup{e},k,0}[n], \mathcal{S}_{0}\right)   + f_1 \mathrm{Av}\left(P_{\textup{e},k,1}[n] , \mathcal{S}_{1}\right).
\label{eq:ErrorProbMAJ-ODD}
\end{equation}

For $N$ being even, the error probability $P_{\textup{e, MAJ},0}^\textup{ even}$ is:
\begin{align}
\begin{split}
 P_{\textup{e, MAJ}, 0}^\textup{ even} =  P_{\textup{e, MAJ}, 0}^\textup{ odd}    -   &\left( \dfrac{f_0}{2}  \mathrm{Av}\left(P_{\textup{e},N/2,0}[n], \mathcal{S}_{0}\right) + \right.  \\   & \hspace{2ex} \left.  +  \dfrac{f_1}{2}   \mathrm{Av}\left(P_{\textup{e},N/2,1}[n] , \mathcal{S}_{1}\right)\right). 
\end{split}
\label{eq:ErrorProbMAJ-EVEN}
\end{align}

If $S=1$ such that $\theta[n]=1$ is associated to the  $x(t_n) \leq x_\textup{th}$, then $P_{\textup{e, OR}, 1} = P_{\textup{e, AND}, 0}$, $P_{\textup{e, AND}, 1} = P_{\textup{e, OR}, 0}$, $P_{\textup{e}, K, 1} = P_{\textup{e}, K, 0}$ and $P_{\textup{e, MAJ}, 1} = P_{\textup{e, MAJ}, 0}$.
\end{theorem}

\begin{IEEEproof}
To compute the average error probability $P_\textup{e}$ given in \eqref{eq:AveErrorProb}, we first need to compute the probability of error events for each decision rule (described in Lemma \ref{lemma:DecisionErrors}), knowing the value of $\theta[n]$. 
We then use the fact that $\theta[n] = 0$ if $x(t_n) \leq x_\textup{th}$ (i.e. $S=0$),  Lemma \ref{lem:prob-K-out-of-N} and Corollary \ref{cor:AND-OR-as-K-out-of-N} to find probabilities for every measurement.
To obtain $P_\textup{e}$, we compute the average error probabilities for $\theta[n] = 0$ or $\theta[n] = 1$ using Definition \ref{def:average} and  their respective frequencies $f_0$ and $f_1$ (Definition \ref{def:defining-theta}).

By De Morgan's law, which says that $\bar{A} + \bar{B} = \overline{A \cdot B}$ for any Boolean variables $A$ and $B$, we find that $P_{\textup{e, OR}, 1} = P_{\textup{e, AND}, 0}$ and $P_{\textup{e, AND}, 1} = P_{\textup{e, OR}, 0}$ when considering $S=1$ is related to $x(t_n) \leq x_\textup{th}$.
The $K$-OUT-OF-$N$ and MAJORITY rules, in turn, are actually independent of how $S$ is assigned so $P_{\textup{e}, K, S=0} = P_{\textup{e}, K, S=1}$ and $P_{\textup{e, MAJ}, 1} = P_{\textup{e, MAJ}, 0}$.
\end{IEEEproof}

\mreview{
\begin{remark}
These results can be written in terms of type I (false-positive) and type II (false-negative) errors.
For $S=0$, the type I error probability is given\footnote{\mreview{The results for MAJORITY is a special case of $K$-OUT-OF-$N$.}}:
\begin{itemize}
\item OR: $\mathrm{Av}\left(1 - P_{\textup{e},N,0}[n] , \mathcal{S}_{0}\right)$,
\vspace{1ex}
\item AND:  $\mathrm{Av}\left(P_{\textup{e},N,0}[n], \mathcal{S}_{0}\right)$,
\item $K$-OUT-OF-$N$: $\sum\limits_{k=K}^{N} \mathrm{Av}\left(P_{\textup{e},k,0}[n], \mathcal{S}_{0}\right)$.
\end{itemize}
Similarly, the type II error probability is given by:
\begin{itemize}
\item OR: $\mathrm{Av}\left(P_{\textup{e},N,1}[n] , \mathcal{S}_{1}\right)$,
\vspace{1ex}
\item AND:  $\mathrm{Av}\left(1 - P_{\textup{e},N,1}[n], \mathcal{S}_{1}\right)$,
\item $K$-OUT-OF-$N$: $\sum\limits_{k=K}^{N} \mathrm{Av}\left(P_{\textup{e},k,1}[n], \mathcal{S}_{1}\right)$.
\end{itemize}
\end{remark}
}

\begin{corollary}
\label{cor:asympt}
The asymptotic behavior of the error probability $P_\textup{e}$ with $N$ for $S=0$ and the different rules is:
\begin{eqnarray}
\lim\limits_{N \rightarrow \infty} P_{\textup{e, OR}, 0} &=&  f_0,\\
\label{eq:ErrorProbOR-asympt}
\lim\limits_{N \rightarrow \infty} P_{\textup{e, AND}, 0} &=&   f_1,\\
\label{eq:ErrorProbAND-asympt}
\lim\limits_{N \rightarrow \infty} P_{\textup{e,}K < \lceil N/2\rceil, 0} &=&  1, \\
\label{eq:ErrorProbK<N/2-asympt}
\lim\limits_{N \rightarrow \infty} P_{\textup{e,}K \geq \lceil N/2\rceil, 0} &=&  0, \\
\label{eq:ErrorProbK>N/2-asympt}
\lim\limits_{N \rightarrow \infty} P_{\textup{e, MAJ},0} &=&  0.
\label{eq:ErrorProbMAJ-asympt}
\end{eqnarray}
\end{corollary}

\mreview{
\begin{remark}
When $N \rightarrow \infty$ and $S=0$, the type I error probability for OR rule tends to $1$, while the type II to $0$.
Conversely, when AND rule is considered, the type I error probability for OR rule tends to $0$, while the type II to $1$.
Therefore, OR always tends to  decide $1$, while AND $0$.
\end{remark}
}

\begin{figure*}[!t]
	\centering
	\includegraphics[width=2\columnwidth]{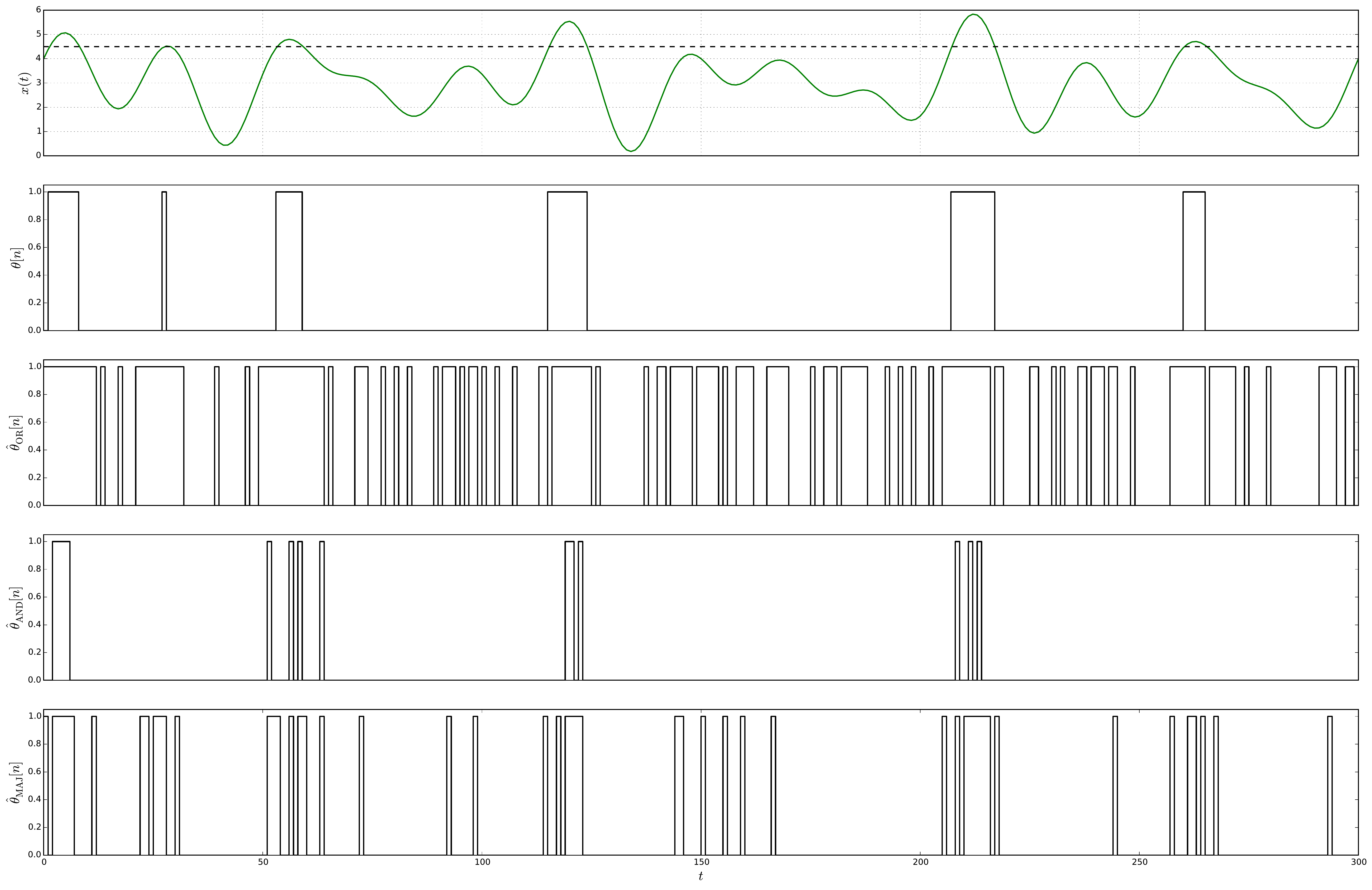}
	\caption{Numerical example of the proposed scenario for the signal $x(t) = \sin(12 \pi t/\eta) + \cos(20 \pi t/\eta) + \sin(26 \pi t/\eta) + 3 $, which has a mean value of $3$ and $\eta=300$, $x_\textup{th} = 4.5$ (represented by the dashed line in the first and third plots, and leading to $\eta_0 = 262$ and $\eta_1 = 38$), $N=3$ (three sensors) and $M=1$ (one hop). Sensor $i$ sets its state $y_i[n]$ at $t_n$ based on the noisy version of signal denoted by $x_i(t_n) = x(t_n) + n_i(t_n)$ where $n_i(t_n)$ is related to an additive Gaussian noise with mean $\mu=0$ and variance $\sigma^2=1$. The error probabilities associated with the sensing procedure are given in \ref{eq:sensing_error1} and \ref{eq:sensing_error2}. After the sensing procedure, signal $y_i[n]$ is sent via a binary symmetric channel with error probability $p_1 = 0.1$, yielding a new signal $s_i[n]$. At the fusion center, a decision is done based on  $s_i[n]$ and the logic operations OR, AND and MAJORITY. Table \ref{tab:example-xth4.5} presents the error probabilities associated with this scenario.}
	\label{fig:3sensors-example}
\end{figure*}

\section{Numerical Analysis}
\label{sec:NumericalAnalysis}
\vspace{-1mm}
\mreview{
To illustrate our framework in a specific setting, let us consider the noisy version of $x(t)$ observed by sensor $i$ such that $x_i(t_n) = x(t_n) + n_i(t_n)$ where $n_i(t_n)$ is related to an additive Gaussian noise with mean $\mu$ and variance $\sigma^2$, defining $y_i[n]$.
Then, the probability $P(x_i(t_n) \leq x_\textup{th})$ that the event $x_i(t_n) \leq x_\textup{th}$ occurs is:
\begin{equation}
P(x_i(t_n) \leq x_\textup{th}) =  
\dfrac{1}{2} \left(1 + \mathrm{erf}\left(\dfrac{x_\textup{th} - x(t_n) - \mu}{ \sigma \sqrt{2}}\right)\right).
\label{eq:sensing_error1}
\end{equation}
Similarly, the probability $P(x_i(t_n) > x_\textup{th})$ is:
\begin{equation}
P(x_i(t_n) > x_\textup{th}) = 
1 - \dfrac{1}{2} \left(1 + \mathrm{erf}\left(\dfrac{x_\textup{th} - x(t_n) - \mu}{ \sigma \sqrt{2}}\right)\right).
\label{eq:sensing_error2}
\end{equation}
}

We study in this section the effect of the number of sensors $N$, the number of hops $M$ and the channel error probability for an input signal $x(t) = \sin(12 \pi t/\eta) + \cos(20 \pi t/\eta) + \sin(26 \pi t/\eta) + 3$ with $\eta=10^4$ and different thresholds $x_\textup{th}$.
We assume that the sensors' observations are affected by  additive Gaussian noise with mean $\mu=0$ and variance $\sigma^2=1$ and all results have been obtained using $S=0$ such that $\theta[n] = 0$ if $x(t_n) \leq x_\textup{th}$.
The results are only presented for OR, AND and MAJORITY rules -- special variations of the $K$-OUT-OF-$N$, as stated in Theorem \ref{theorem:ErrorProbDecision} and Corollary \ref{cor:AND-OR-as-K-out-of-N}.
It is worth mentioning that, although this setting is somehow arbitrary, the analytic framework proposed here can be applied to different scenarios; the computational experiments presented in this section are coded in python language using IPython framework and are available at \cite{Simulator}.

Before starting, we would like to explain our choice of the input signal $x(t)$.
Our idea was to have a positive deterministic signal, limited in amplitude, that has different peaks and a visually ``interesting'' behavior.
This allows us to see non-linear effects of changes in $x_\textup{th}$.
Our choice $x(t) = \sin(12 \pi t/\eta) + \cos(20 \pi t/\eta) + \sin(26 \pi t/\eta) + 3$, whose amplitude varies from $0$ to $6$ is illustrated in the first plot of Fig. \ref{fig:3sensors-example}, but for $\eta=300$ and assuming $M=1$ and $p_1=0.1$, $N=3$ with additive Gaussian noise ($\mu=0$, $\sigma^2=1$). 

The top plot represents the signal $x(t)$ while the actual system state $\theta[n]$ associated with the event $x(t_n) > x_\mathrm{th}$ is shown next. 
We present in the third plot the estimations $x_i(t_n)$ from the three sensors based on their noisy version of $x(t)$, followed by their respective states $y_i[n]$.
The three received signals $s_i[n]$ at the fusion center, after passing through the communication link, are presented next. 
The last plots represent the decisions using OR, AND and MAJORITY.

\begin{table}[!b]
\vspace{-4mm}
	\centering
	\mreview{
	\caption{Error probability for the snapshot presented in Fig. \ref{fig:3sensors-example}}
	\label{tab:example-xth4.5}	
	\centering
	\begin{tabular}{|c|c|c|c|c|}
		\hline
		Decision rule  &	Analytic  & \multicolumn{2}{c|}{Simulation} \\ \hline
		$g$	& $P_\textup{e}$ & $P_\textup{e}$ ($\eta=300$) & $P_\textup{e}$ ($\eta=10^{5}$) \\	\hline \hline
		OR 				  & 	$0.383$      & $0.366$	& $0.381$	\\ \hline
		AND 			  & 	$0.101$      & $0.090$  & $0.101$	\\ \hline
		MAJORITY 		  & 	$0.129$      & $0.136$	& $0.125$	\\\hline	
	\end{tabular}
	\vspace{-4mm}
	}
\end{table}

Table~\ref{tab:example-xth4.5} compares the simulated and analytic error probabilities for this example, which for better visualization only considers $\eta=300$. Notice that for this particular example AND rule has better performance compared with the other two. As we shall see later, AND rule leads to small error probabilities when the number of sensors is low.  

\begin{figure}[t]
	\centering
	\includegraphics[width=\columnwidth]{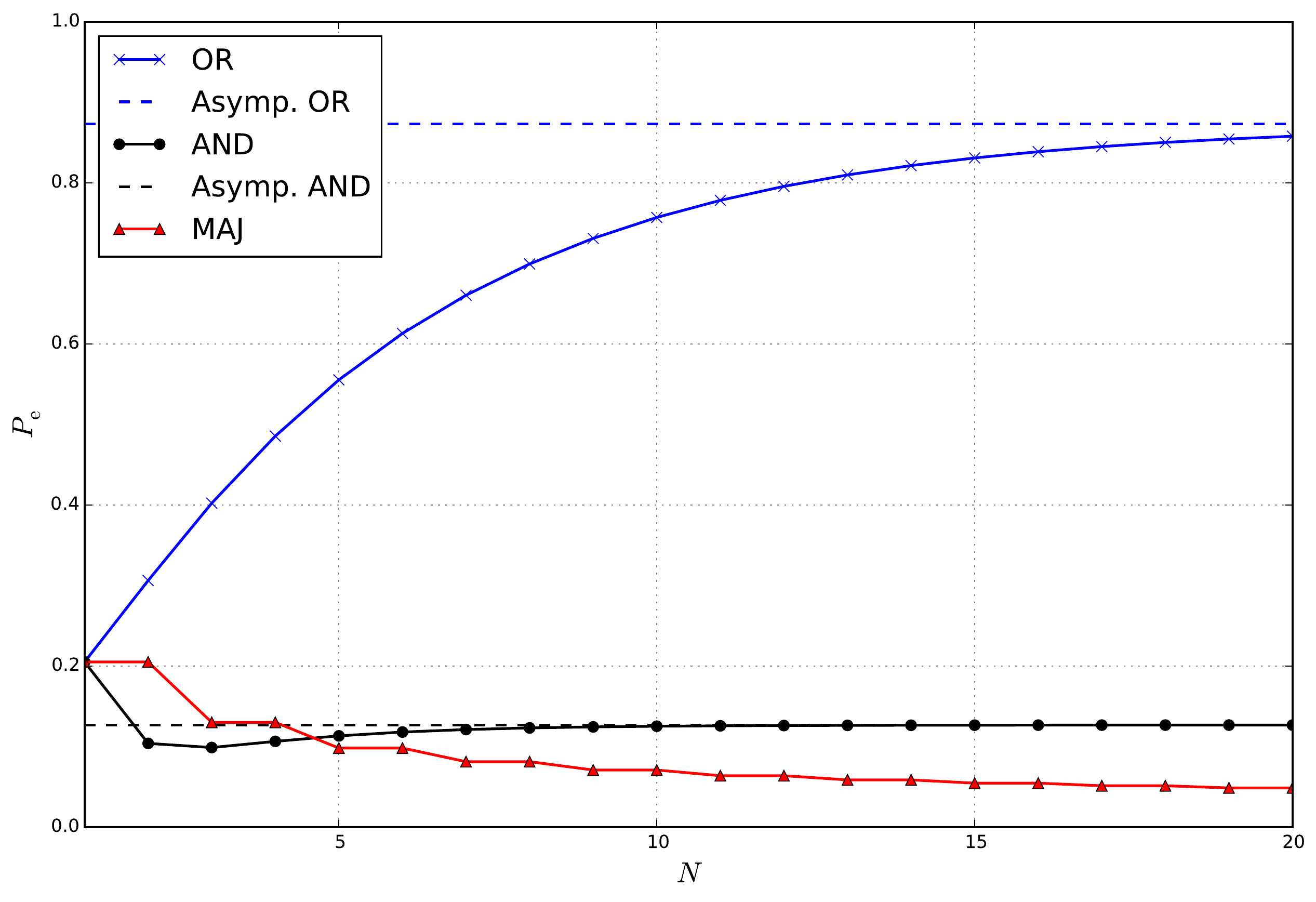}
	\caption{Average error probability $P_\textup{e}$ given in Theorem \ref{theorem:ErrorProbDecision} as a function of the number of sensors $N$ for OR, AND and MAJORITY decision rules, threshold $x_\textup{th}=4.5$, $M=1$ and $p_1=0.1$. The total number of samples is $\eta=10^4$ where $\eta_0 = 8747$ and $\eta_1 = 1253$.}
	\label{fig:Pe-vs-N-xth-4-5}
\vspace{-3mm}
\end{figure} 

\begin{figure}[!t]
	\centering
	\includegraphics[width=\columnwidth]{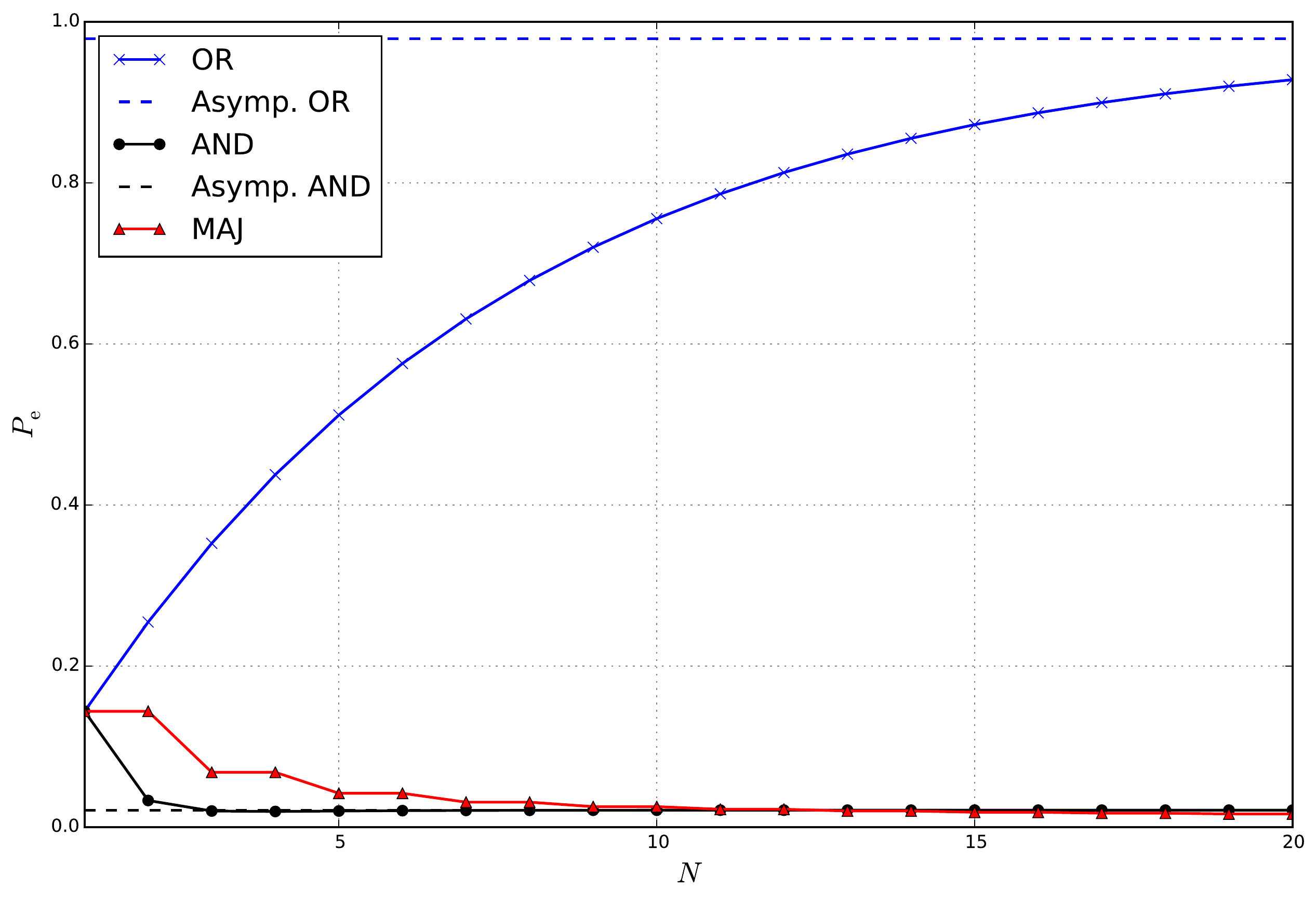}
	\caption{Average error probability $P_\textup{e}$ given in Theorem \ref{theorem:ErrorProbDecision} as a function of the number of sensors $N$ for OR, AND and MAJORITY decision rules, threshold $x_\textup{th}=5.5$, $M=1$ and $p_1=0.1$. The total number of samples is $\eta=10^4$ where $\eta_0 = 9790$ and $\eta_1 = 210$.}
	\label{fig:Pe-vs-N-xth-5-5}
\vspace{-5mm}
\end{figure}

Fig. \ref{fig:Pe-vs-N-xth-4-5} illustrates how the average error probability $P_\textup{e}$ using OR, AND and MAJORITY rules varies with the number $N$ of sensors making measurements when the threshold that defines the event $\mathcal{E}$ is $x_\textup{th}=4.5$.
The plot shows that: ($i$) OR rule has a high error probability, which increases with $N$, ($ii$) AND rule has the best performance for small values of $N$, and ($iii$) MAJORITY is the best choice when $N$ grows such that $P_\textup{e}$ tends to $0$.
To analyze these facts, we need to deal with $\theta[n]$. 
From Theorem \ref{theorem:ErrorProbDecision} and Corollary \ref{cor:asympt}, the frequency of $\theta[n]=0$ and $\theta[n]=1$ determines the performance of OR and AND so the former outperforms when $\theta[n]=0$  is more frequent, and vice-versa.
In our example, the frequencies are: $f_0 = 0.8747 $ and $f_1=0.1253$, providing their asymptotic limit.

For MAJORITY, the asymptotic performance is independent of such frequencies.
When a small number of sensors is considered, however, it does not provide the best performance since an error in more than $N/2$ signal is not rare.
In this case, AND is the best, even working below its asymptotic limit.
This happens due to the way that the logic operation AND works, balancing the error events when $\theta[n]=0$ and $\theta[n]=1$.

\begin{figure}[!t]
	\centering
	\includegraphics[width=\columnwidth]{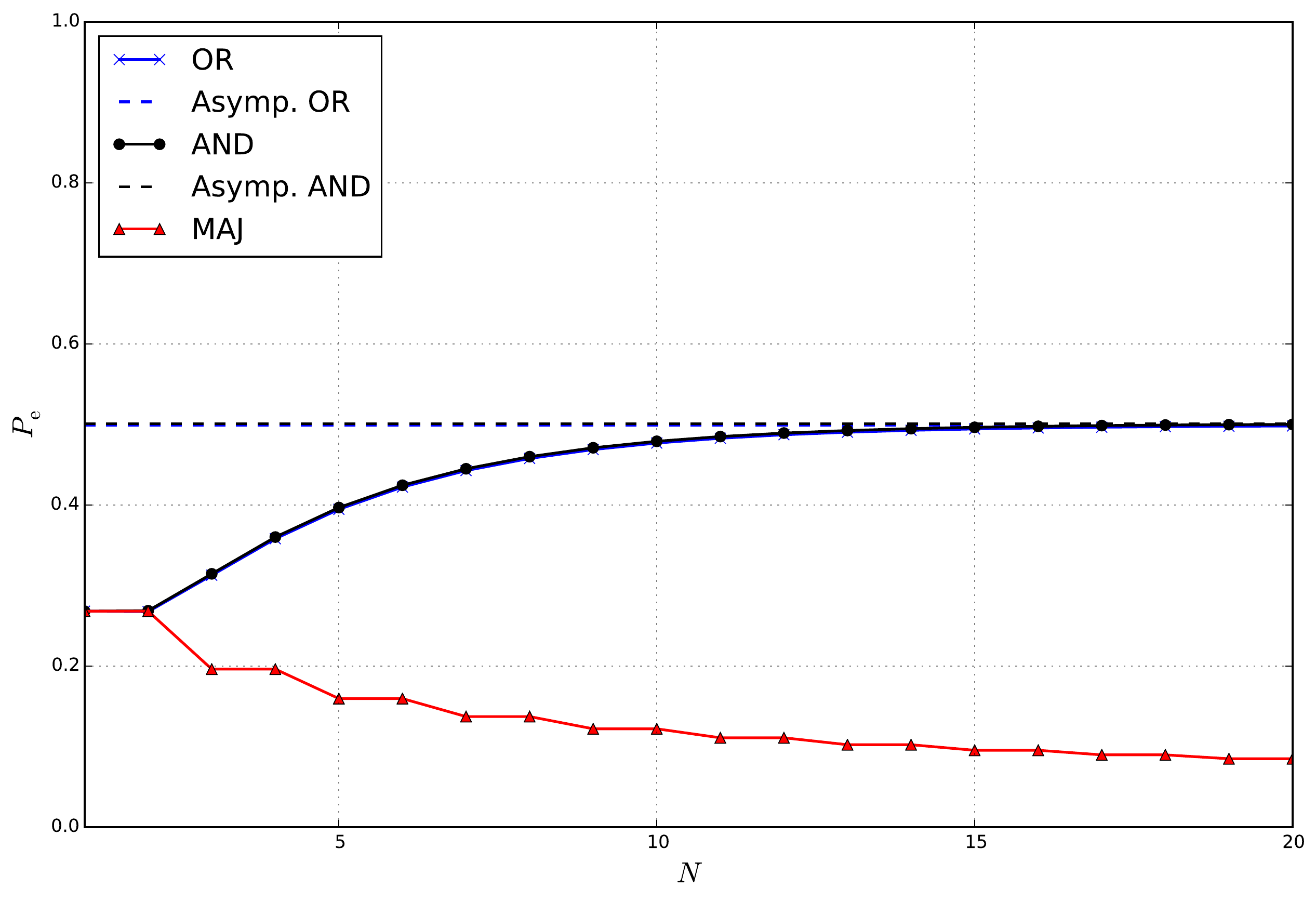}
	\caption{Average error probability $P_\textup{e}$ given in Theorem \ref{theorem:ErrorProbDecision} as a function of the number of sensors $N$ for OR, AND and MAJORITY decision rules, threshold $x_\textup{th}=3$, $M=1$ and $p_1=0.1$. The total number of samples is $\eta=10^4$ where $\eta_0 = 4992$ and $\eta_1 = 5008$.}
	\label{fig:Pe-vs-N-xth-3}
	\vspace{-4mm}
\end{figure}

\begin{figure}[!t]
	\centering
	\includegraphics[width=\columnwidth]{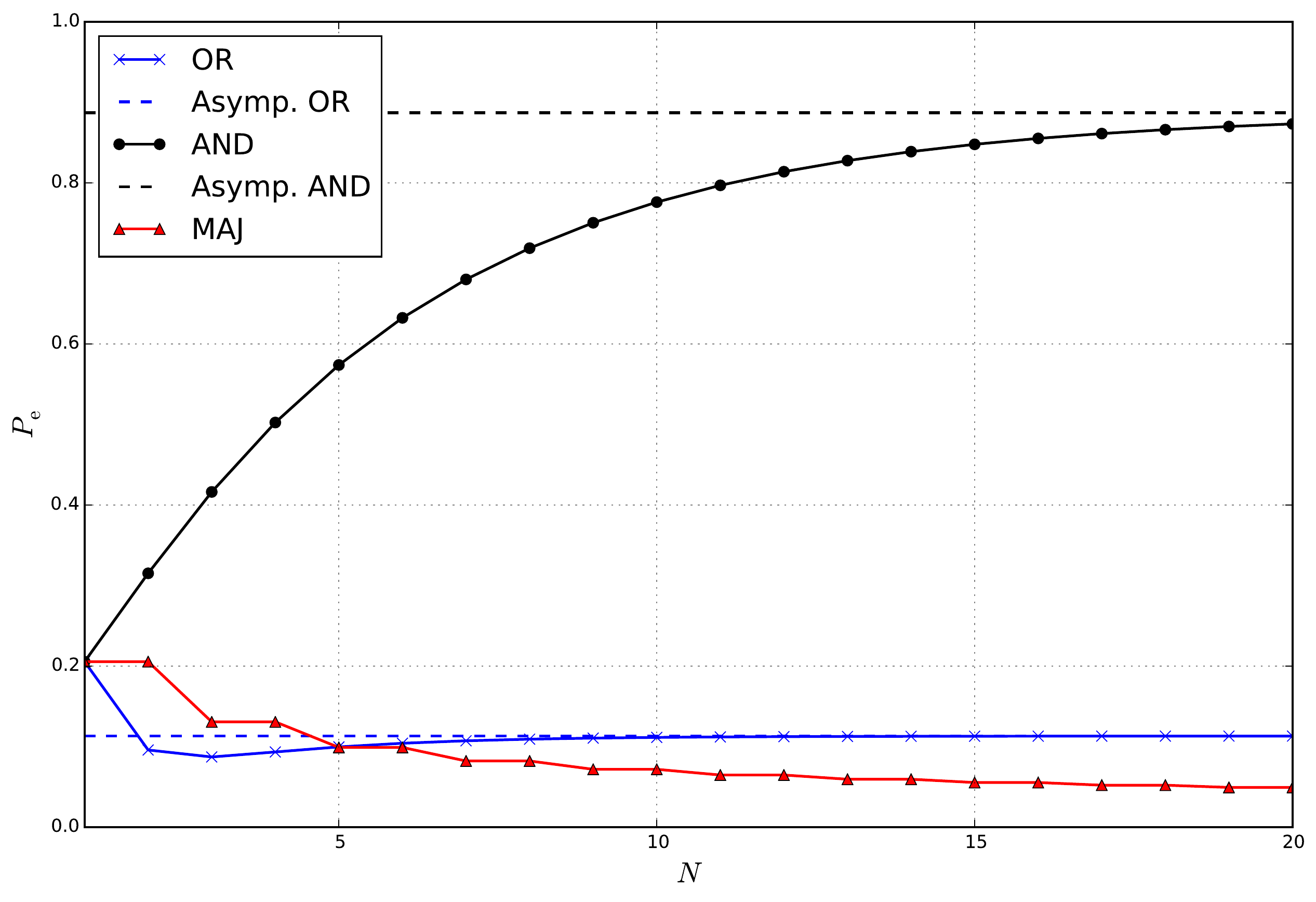}
	\caption{Average error probability $P_\textup{e}$ given in Theorem \ref{theorem:ErrorProbDecision} as a function of the number of sensors $N$ for OR, AND and MAJORITY decision rules, threshold $x_\textup{th}=1.5$, $M=1$ and $p_1=0.1$. The total number of samples is $\eta=10^4$ where $\eta_0 = 1128$ and $\eta_1 = 8872$.}
	\label{fig:Pe-vs-N-xth-1-5}
\vspace{-4mm}
\end{figure}

To get more insights on the system performance, Figs. \ref{fig:Pe-vs-N-xth-5-5},  \ref{fig:Pe-vs-N-xth-3} and \ref{fig:Pe-vs-N-xth-1-5} show the average error probability as a function $N$ for $x_\textup{th}=5.5$ ($f_0=0.979$ and $f_1=0.021$), $x_\textup{th}=3$ ($f_0=0.4992$ and $f_1=0.5008$) and $x_\textup{th}=1.5$ ($f_0=0.1128$ and $f_1=0.8872$), respectively.
In the scenario where $x_\textup{th}=5.5$, a similar behavior to the $x_\textup{th}=4.5$ is observed, but with the AND rule having a better performance due to the smaller frequency $f_1$ of events $\theta[n]=1$.
On the other hand, when $x_\textup{th}=1.5$, the performance of the OR and AND rules switches  in relation to when $x_\textup{th}=4.5$ as far as the frequencies $f_0$ and $f_1$ have also switched; now OR works better because $\theta[n]=0$ is much more frequent.
When the $\theta[n]$ is more evenly distributed, illustrated in the scenario where $x_\textup{th}=3$, AND and OR are equivalent and their error probability tends to $0.5$ (which is basically a random guess of the input state).
In all scenarios, the MAJORITY rule maintains its asymptotic optimal performance, working better and better when the number of sensors grows.


\begin{figure}[!t]
	\centering
	\includegraphics[width=\columnwidth]{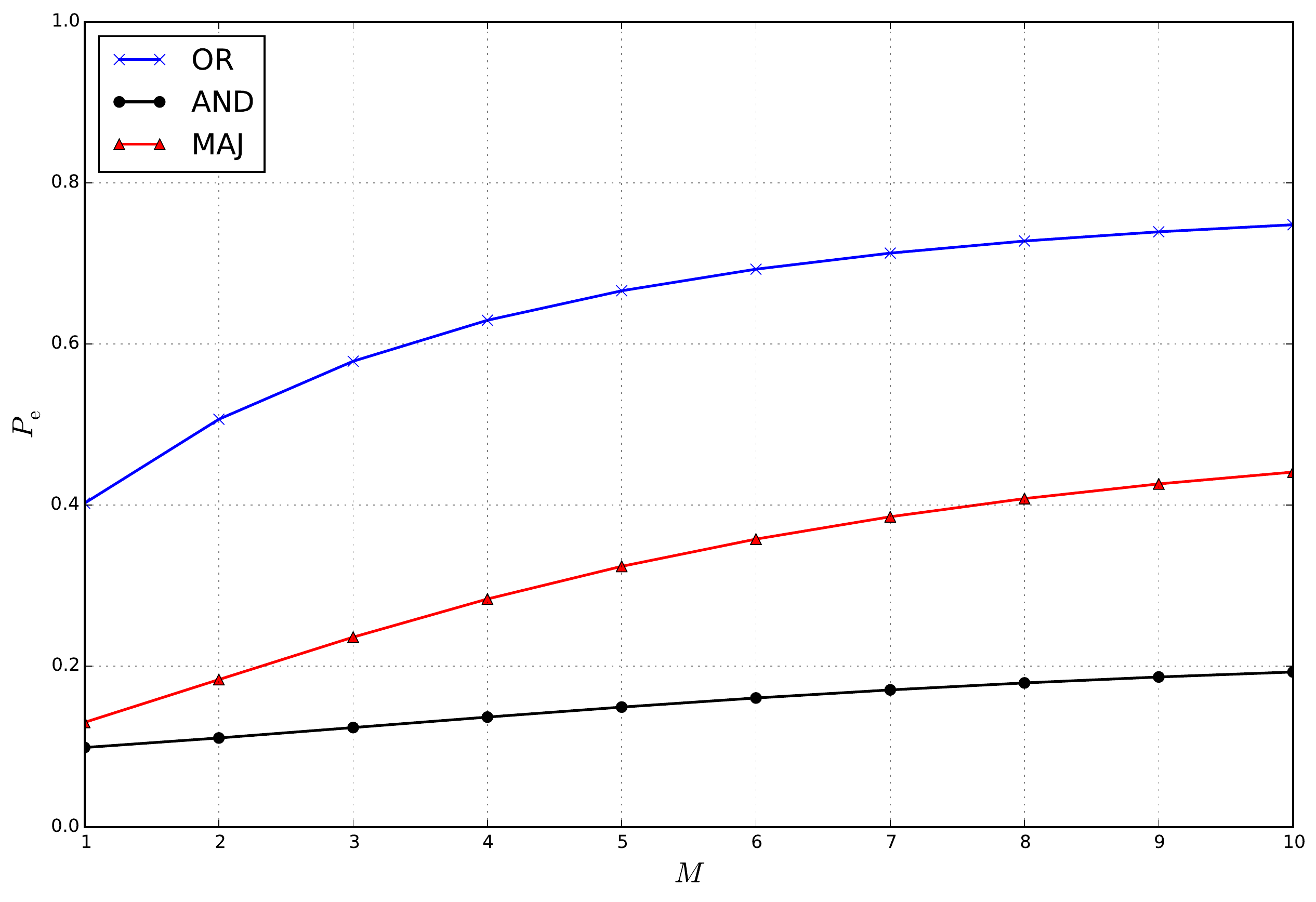}
	\caption{Average error probability $P_\textup{e}$ given in Theorem \ref{theorem:ErrorProbDecision} as a function of the number of hops $M$ considering $p_j=p=0.1$ for OR, AND and MAJORITY rules, threshold $x_\textup{th}=4.5$ and $N=3$. The equivalent channel probability after $M$ hops is given in Corollary \ref{cor:symmetric-BSC-cascade}.}
	\label{fig:Pe-vs-M-xth-4-5}
\vspace{-5mm}
\end{figure}

Fig. \ref{fig:Pe-vs-M-xth-4-5} presents the effects of the number of hops on the average error probability for $x_\textup{th}=4.5$ and $N=3$, considering that every one of the $M$ hops is modeled as a binary symmetric channel with the same error probability $p=0.1$ so that the equivalent channel has the error probability given by Corollary \ref{cor:symmetric-BSC-cascade}.
As one would expect, the increase of the number of hops $M$ also increases the average error probability, regardless of the decision rule.
For the setting considered here, the AND rule seems more robust against the increase of $M$, so that the error probability associated to it grows slower than the other two options.
Once again, this fact occurs due to the way AND  balances the error probabilities for the two possible values of $\theta[n]$, given more weight (in proportion to its occurrence) to less frequent error events.
The OR rule, on the other hand, has an overall poor performance because it balances the error probability in the opposite way, which leads to even worse error probabilities.
The MAJORITY rule appears to be more susceptible to the increase of $M$ than AND, which indicates that the increase of the equivalent channel error probability when $M$ grows seems to affect more the former.

\begin{figure}[!t]
	\centering
	\includegraphics[width=\columnwidth]{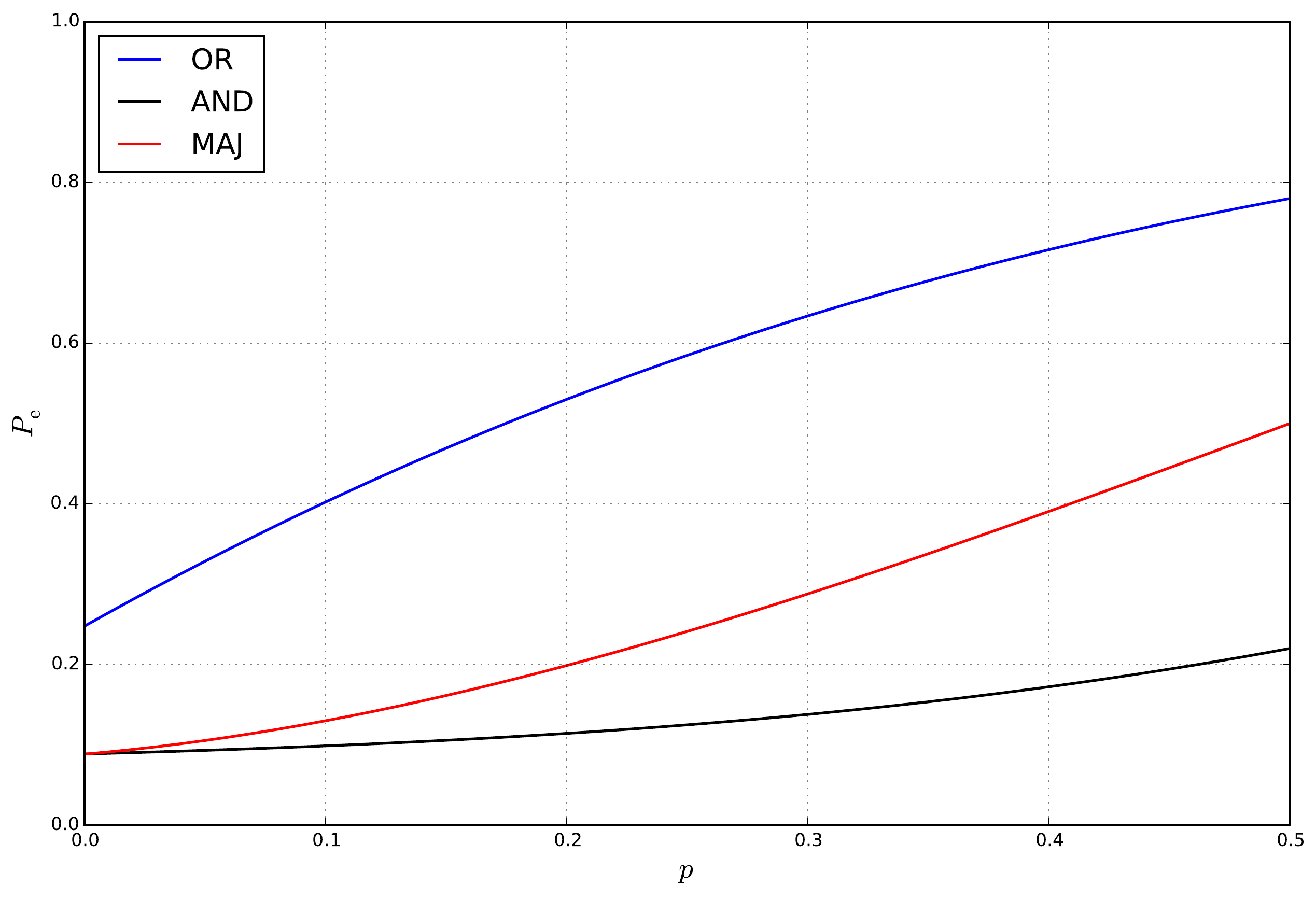}
	\caption{Average error probability $P_\textup{e}$ given in Theorem \ref{theorem:ErrorProbDecision} as a function of channel error probability $p_1$ considering only $1$-hop transmission (i.e. $M=1$) for OR, AND and MAJORITY rules, threshold $x_\textup{th}=4.5$ and $N=3$.}
	\label{fig:Pe-vs-p-xth-4-5}
\vspace{-5mm}
\end{figure}

To better understand this fact, we present in Fig. \ref{fig:Pe-vs-p-xth-4-5} how the average error probability varies with the channel error probability $p_1$ when $x_\textup{th}=4.5$, $M=1$ and $N=3$.
We now can see clearer that the MAJORITY rule is indeed more susceptible to worse channel conditions than the AND rule for the scenario under analysis.
As presented in Theorem \ref{theorem:ErrorProbDecision}, MAJORITY, different from the latter rule, does not favor the error events in different ways so that when the channel error probability increases, it will increase in the same proportion for both the more and less frequent states.
Consequently, although MAJORITY asymptoticly outperforms the others in terms of the number of sensors, it is much more vulnerable to an increase of the channel error probabilities.

\vspace{-3mm}
\section{Final remarks}
\label{sec:Discussions-final-remarks}
\vspace{-1mm}
In this paper, we analyzed different ways that one could design a relatively simple WSN based on three phases, namely sensing, communication and decision.
Different from the literature of distributed sensing and estimation, we targeted at implementing simple decision rules, regardless of their optimality. 
Our idea here was to show that it is possible to attain low error probabilities using a simple threshold based quantizer, a limited bandwidth of $1$ bit and low complexity decision rules such as AND, OR and MAJORITY.

If the occurrence of the event is rare and associated with the state ``$1$'', the AND rule can lead to a low error probability with a small number of sensors, although MAJORITY can asymptotically reach $0$-error probability for a large the number of sensors.
If its occurrence is more equally distributed, then AND and OR rules have a poor performance while the MAJORITY is better than the other options but still requiring a relatively large number of sensors.
We also show that the MAJORITY rule is more susceptible to  channel errors than the AND rule, reflecting the way that it balances the error events with the input state frequencies.

In any case, our results indicate a simple and cheap way to implement a WSN when the application does not have strict requirements.
\mreview{
Our plan is to extend these results by considering erasure channels as in \cite{zhang2013detection} and advanced relay strategies as in \cite{hirlet-2014,carliche-2016}, assuming the nodes follow a specific spatial distribution as in \cite{zhang2015event}.
Sensors experiencing different input signals (e.g. smart-metering or sensors in different rooms) also constitute an interesting extension of the present work.
Other promising direction is to individually analyze the types I and II error probabilities such keeping one of them as a fixed target (e.g. applications that require a very low false negative probability, while false positives are unconstrained).
}

\bibliographystyle{IEEEtran}

%
\end{document}